\documentclass[12pt]{article}
\usepackage[cp1251]{inputenc}
\usepackage[english]{babel}
\usepackage{mathtext,amsfonts,euscript,graphicx,amsmath,amssymb,color}
\usepackage[superscript,ref,biblabel]{cite}
\newcommand {\Rs}{R_{\sigma}}
\newcommand {\ninf}{n_{\infty}}

\begin{document}
\selectlanguage{english} 

\title{Dynamics of gas bubble growth \\ in~a~supersaturated solution with \\ Sievert's solubility law}
\author{G. Yu. Gor, A. E. Kuchma}

\maketitle

\begin{center}
\textit{Department of Theoretical Physics, Research Institute of Physics,\\
Saint-Petersburg State University,\\
1 Ulyanovskaya Street, Petrodvorets, St. Petersburg, 198504, Russia}
\end{center}

\begin{abstract}
\noindent
This paper presents a theoretical description of diffusion growth of a gas bubble after its nucleation in supersaturated liquid solution. We study systems where gas molecules completely dissociate in the solvent into two parts, thus making Sievert's solubility law valid. We show that the difference between Henry's and Sievert's laws for chemical equilibrium conditions causes the difference in bubble growth dynamics. Assuming that diffusion flux is steady we obtain a differential equation on bubble radius. Bubble dynamics equation is solved analytically for the case of homogeneous nucleation of a bubble, which takes place at a significant pressure drop. We also obtain conditions of diffusion flux steadiness. The fulfillment of these conditions is studied for the case of nucleation of water vapor bubbles in magmatic melts.
\end{abstract}

\newpage

\section{Introduction}
\label{sec:Introduction}

Nucleation and growth of gas bubble in gas-liquid solutions are highly widespread phenomena both in nature and in technological processes. Gas bubble growth plays an important role in the production of microcellular materials and foams, as well as in glass production \cite{Cable-Frade}. Besides, the growth of water vapor bubbles dissolved in magmatic melt is a process which leads to volcanic eruptions \cite{Sahagian-1999, Navon, Lensky, Chernov-2004}.

This paper presents a theoretical description of diffusion growth of a gas bubble in liquid solution resulting from a pressure drop (i.~e. instant creation of supersaturation). It has to be noted that growth regularities of a solitary new phase embryo are crucial for the description of the evolution of the whole process of the phase transition in supersaturated solution \cite{Chernov-2004, Slezov-2004, Slezov-2005, Lifshitz-Slezov-1961}.
 
The description of gas bubble growth in supersaturated solution traditionally employs two rough approximations \cite{Navon, Lensky, Chernov-2004, Slezov-2004, Slezov-2005}: the flux of the dissolved gas toward the bubble is assumed to be steady; and the consideration is limited only to bubbles with such a large radius that it is possible to neglect for Laplace pressure in comparison with the external pressure of the solution.

In the present paper, while exploiting the steady approximation, considering the bubble from the very moment of its nucleation, we take into account time-dependent Laplace pressure in the bubble. This approach makes both gas density in the bubble and the equilibrium concentration of the dissolved gas at the surface of the bubble time-dependent as well (hereinafter we will refer to number density of molecules using term "concentration").

Gas bubble growth in a solution taking into account Laplace pressure was considered as early as in 1950 in the classical paper by Epstein and Plesset \cite{Epstein-Plesset}, where the authors obtained the equation for the bubble radius as a function of time. In order to correlate the equilibrium concentration of the dissolved gas and the solution pressure, Ref. \cite{Epstein-Plesset} presupposed the fulfillment of Henry's law: i. e. the proportional dependence between these two values. Indeed, Henry's law is fulfilled in many systems, e. g. in case of the solution of $\text{CO}_2$ in magmatic melt \cite{Lensky}; however, it is not universal: one cannot apply it to the case of the solution of $\text{H}_2\text{O}$ vapor \cite{Chernov-2004, Navon}. In the latter case, which is crucial for practical reasons, Sievert's law is observed: the equilibrium concentration of the dissolved gas is proportional to the square root of the solution pressure.

In the current paper we will obtain the equation for the bubble radius as a function of time similarly to the equation obtained in Ref. \cite{Epstein-Plesset} for the case of Henry's law. We will show that the difference between Henry's and Sievert's laws for chemical equilibrium conditions at the bubble surface causes the difference in bubble growth dynamics. We will obtain conditions when the steady approximation for the diffusion flux of gas molecules toward the bubble is applicable. We will show that, as the radius of the bubble increases, the steady condition becomes stricter; and, consequently, as a rule, the steady regime of bubble growth gradually gives way to the nonsteady one.

The case of homogeneous nucleation of the bubble, which takes place at the significant pressure drop, will be supplied with the analytical solution for the obtained bubble dynamics equation.

In section \ref{sec:Volcanoes} of the current paper we will analyze whether the steady approximation is applicable to the case of gas bubbles in magma described by Navon \cite{Navon} and Chernov et al. \cite{Chernov-2004} for large radius of a bubble (neglecting Laplace forces).

\section{Equilibrium concentration of dissolved gas: Henry's law and Sievert's law}
\label{sec:Equilibrium}

Let us consider a gas solution in liquid which previously was in equilibrium state at temperature $T$ and pressure $P_0$. The concentration of the dissolved gas in the solution under such conditions will be denoted as $n_0$.

Then we instantly relieve the external pressure to value $\Pi$ in such a way that solution becomes supersaturated. The temperature and the volume of the solution remain the same, thus value $n_0$ still serves for the dissolved gas concentration.

It is more convenient to express the state of the solution in terms of dimensionless variables: supersaturation $\zeta$ and gas solubility $s$
\begin{equation}
\label{zeta}
\zeta \equiv \frac{n_0 - \ninf}{\ninf},
\end{equation}
\begin{equation}
\label{s}
s \equiv \frac{k T \ninf}{\Pi},
\end{equation}
where $n_{\infty}$ is the equilibrium concentration of dissolved gas, $k$ is the Boltzmann's constant.

Some time after the pressure drop a gas bubble nucleates and starts growing regularly. We assume that the bubble is in mechanical equilibrium with the solution, and its dynamics is governed only by diffusion process. We will not discuss the settlement of the mechanical equilibrium between the bubble and the solution, assuming that this settlement occurs much faster than the settlement of chemical equilibrium, as in Ref. \cite{Chernov-2004}. The time of the bubble nucleation is considered as $t = 0$. The radius of the bubble will be denoted as $R$. Following \cite{Cable-Frade, Sahagian-1999, Navon, Lensky, Chernov-2004, Slezov-2004, Slezov-2005} we treat the liquid solvent as non-volatile (which is true for systems considered in section \ref{sec:Volcanoes}); therefore, we assume that the bubble consists of gas only, but not of the solvent vapor. The gas in the bubble is considered to be ideal.

When the bubble is studied after sufficiently long time since its nucleation, its radius can comply with the strong inequality
\begin{equation}
\label{R>>Laplace}
R \gg 2 \sigma /\Pi,
\end{equation}
where $\sigma$ is the surface tension of the pure solvent (it is true while the solution is considered diluted). Eq. \eqref{R>>Laplace} makes it possible to neglect the influence of Laplace forces on the bubble growth. Therefore, we can write the following equations for the pressure in the bubble $P_R$ and for the equilibrium solution concentration near the surface of the bubble $n_R$:
\begin{equation}
\label{P=Pi}
P_R = \Pi,
\end{equation}
\begin{equation}
\label{n_R=n_inf}
n_R = n_{\infty}.
\end{equation}
The subscription $\infty$ denotes that the equilibrium concentration $\ninf$ is related with the equilibrium near the flat surface of phase separation ($R \rightarrow \infty$). As follows from Eq. \eqref{P=Pi}, the gas concentration in the bubble $n_g$ is constant. Using the ideal gas law we have
\begin{equation}
\label{n_g}
n_g = \frac{\Pi}{k T}.
\end{equation}

When Eqs. \eqref{R>>Laplace}, \eqref{P=Pi}, \eqref{n_R=n_inf} are fulfilled, the description of bubble dynamics is trivial for the case of steady-state diffusion and can be described analytically even for the non-steady case \cite{Scriven-1959, Grinin-Kuni-Gor}.

From the moment of bubble nucleation and till Eq. \eqref{R>>Laplace} becomes valid, Laplace forces influence bubble growth. Thus quantities $P_R$ and $n_R$ become radius-dependent (and, therefore, time-dependent). So, for $P_R$ we now have
\begin{equation}
\label{Mech-Eq}
P_R = \Pi + \frac{2 \sigma}{R}.
\end{equation}

In order to write the equation for $n_R$ we need to know the solubility law. For the simplest case of Henry's law the equilibrium concentration of dissolved gas is proportional to the corresponding pressure
\begin{equation}
\label{Chem-Eq-Henry}
\frac{n_R}{\ninf} = \frac{P_R}{\Pi}.
\end{equation}
The study of bubble dynamics with consideration of Laplace forces for the case of Henry's law was discussed in \cite{Epstein-Plesset, Kuchma-Gor-Kuni} . However, Henry's law is valid only in systems where there no molecules dissociation occurs during gas dissolution. In another important case, when the gas molecules completely dissociate in the solvent into two parts, Henry's law is replaced with so-called Sievert's law \cite{Darken-Gurry, Landau-V} , and, therefore, Eq. \eqref{Chem-Eq-Henry} is replaced with the following one:
\begin{equation}
\label{Chem-Eq}
\frac{n_R}{\ninf} = \sqrt{\frac{P_R}{\Pi}}.
\end{equation}
Sievert's law is fulfilled for the case of water vapor dissolved in a silicate melt \cite{Stolper}. The study of such solutions is important both for glass production \cite{Cable-Frade} and for volcanic systems \cite{Navon, Chernov-2004} .

The replacement of Eq. \eqref{Chem-Eq-Henry} with Eq. \eqref{Chem-Eq} means that the boundary condition for the gas diffusion problem will be different for Henry's and Sievert's laws. The change of boundary condition, as we will see further, leads to the change in bubble dynamics.

\section{Bubble dynamics equation}
\label{sec:Dynamics}

After the nucleation of the bubble, when its growth can be considered regular (i. e. the bubble can not be dissolved by fluctuations), its growth is governed by the diffusion flux of gas molecules into it. In this paper we will study the case when the diffusion flux can be assumed as steady. The conditions when such approximation is valid will be discussed in section \ref{sec:Stationarity}.

Considering the equality of gas concentration at the bubble surface and the equilibrium concentration $n_R$ and the equality of gas concentration far from the bubble and the initial concentration $n_0$, we can derive a simple expression for the steady flux density $j_D$ 
\begin{equation}
\label{Flux}
j_D = D \frac{n_0 - n_R}{R}.
\end{equation}
Here $D$ is the diffusion coefficient of gas molecules in pure solvent (we assume that the solution is diluted).

Now let us write the expression for the number of gas molecules $N$ in the bubble. Exploiting the ideal gas law and using Eq. \eqref{Mech-Eq} we have
\begin{equation}
\label{dN/dt-simple}
N = \frac{4 \pi}{3 k T} R^3 \left[ \Pi + \frac{2 \sigma}{R}\right].
\end{equation}
Differentiating Eq. \eqref{dN/dt-simple}, using denotation \eqref{n_g}, we obtain
\begin{equation}
\label{dN/dt}
\frac{dN}{dt} = 4 \pi n_g R^2 \frac{dR}{dt} \left[ 1 + \frac{\Rs}{R} \right],
\end{equation}
where
\begin{equation}
\label{R_s}
\Rs \equiv \frac{4}{3} \frac{\sigma}{\Pi}
\end{equation}
is the characteristic size of the bubble.

The material balance between the dissolved gas and the gas in the growing bubble gives us the following equality
\begin{equation}
\label{Balance}
\frac{dN}{dt} = 4 \pi R^2 j_D.
\end{equation}
Substituting expression for diffusion flux density \eqref{Flux} and the rate of change of the number of molecules Eq. \eqref{dN/dt} into material balance equation \eqref{Balance} and then, exploiting Eqs. \eqref{s} and \eqref{n_g}, we have
\begin{equation}
\label{Dyn-Eq-2}
R \dot{R} \left[ 1 + \frac{\Rs}{R} \right] = D s \frac{n_0 - n_R}{\ninf}.
\end{equation}
Using Eqs. \eqref{zeta}, \eqref{Mech-Eq}, \eqref{Chem-Eq} and \eqref{R_s}, the fraction in the r. h. s. of Eq. \eqref{Dyn-Eq-2} can be expressed as
\begin{equation}
\label{Conc-Frac}
\frac{n_0 - n_R}{\ninf} = (\zeta + 1) - \sqrt{1 + \frac{3}{2} \frac{\Rs}{R}}.
\end{equation}
Eq. \eqref{Conc-Frac} allows us to rewrite the equation of bubble dynamics \eqref{Dyn-Eq-2} in its final form: 
\begin{equation}
\label{Dyn-Eq}
R \dot{R} \left[ 1 + \frac{\Rs}{R} \right] = D s \left[ (\zeta + 1) - \sqrt{1 + \frac{3}{2} \frac{\Rs}{R}} \right].
\end{equation}

It should be noted that the well-known paper by Cable and Frade presented only a special case of Eq. \eqref{Dyn-Eq}, when $\zeta = -1$ (Eq. (34) in Ref. \cite{Cable-Frade}) ; such value of supersaturation corresponds to the dissolution of gas bubble in the pure solvent. The form of this equation for the general case presented in the current paper has not been obtained previously.

\section{Fluctuational nucleation, critical bubble and initial condition for the bubble growth}
\label{sec:Critical}

All aforesaid (sections \ref{sec:Introduction}, \ref{sec:Equilibrium}, \ref{sec:Dynamics} and the obtained equation for the bubble growth dynamics \eqref{Dyn-Eq}) can be applied to the cases of both homogeneous and heterogeneous nucleation. Below we will consider only the homogeneous nucleation case. 

Homogeneous nucleation of a gas bubble in a supersaturation solution means fluctuational mechanism of its appearance and requires a significant pressure drop (of the order of $10^3$ times). It is these conditions of bubble nucleation that exist in magmatic melts during volcanic eruptions \cite{Chernov-2004} . Since we decided to describe only regular growth of a bubble, we need to exclude the very process of nucleation from our examination and consider a bubble when it is already supercritical.

Under the notion of critical bubble we traditionally \cite{Skripov-1972, Slezov-2004} understand such a bubble which radius $R_c$ corresponds to the extremum of work of bubble formation. A critical bubble is in mechanical equilibrium with solution at the initial pressure $P_0$ and in chemical equilibrium with solution with concentration $n_0$. These two conditions together with the solubility law unambiguously define the value of $R_c$.
 
The mechanical equilibrium condition, with Eq. \eqref{Mech-Eq}, evidently gives us
\begin{equation}
\label{R_c-def}
R_c = \frac{2 \sigma}{P_0 - \Pi},
\end{equation}
and chemical equilibrium condition allows us to relate value $P_0$ to the supersaturation $\zeta$.

From Eq. \eqref{Chem-Eq} we have
\begin{equation}
\label{n_0/n_inf}
\frac{n_0}{\ninf} = \sqrt{\frac{P_0}{\Pi}},
\end{equation}
or, using Eq. \eqref{zeta}, 
\begin{equation}
\label{P_0-zeta}
P_0 = \Pi \left( \zeta + 1 \right)^2.
\end{equation}
Finally, substituting Eq. \eqref{P_0-zeta} in Eq. \eqref{R_c-def} for the radius of a critical bubble, we have
\begin{equation}
\label{R_c}
R_c = \frac{2 \sigma}{\Pi \left[ (\zeta + 1)^2 - 1 \right]}.
\end{equation}
Using Eq. \eqref{R_c} and Eq. \eqref{R_s} we can write the relation between $\Rs$ and $R_c$
\begin{equation}
\label{R_s-R_c}
\Rs = \frac{2}{3} \left[ (\zeta + 1)^2 - 1 \right] R_c
\end{equation}
which will be exploited later.

Here it is important to make a remark considering quantity $s$ -- gas solubility. Eqs. \eqref{zeta} and \eqref{s} give us
\begin{equation}
\label{s(zeta)}
s = \frac{n_0 k T}{\Pi} \frac{1}{\zeta + 1}.
\end{equation}
If Henry's law was fulfilled, we would have $\zeta + 1 = P_0/\Pi$ and, therefore,
\begin{equation}
\label{s-Henry}
s = \frac{n_0 k T}{P_0}, 
\end{equation}
$s$ would be a tabular value defined only by the initial state of the solution (before the pressure drop).

Since here we consider Sievert's law, we have (see Eq. \eqref{P_0-zeta}) $\zeta + 1 = \sqrt{P_0/\Pi}$; thus instead of Eq. \eqref{s-Henry}, we obtain
\begin{equation}
\label{s-Sievert}
s = \frac{n_0 k T}{\sqrt{P_0 \Pi}}.
\end{equation}
Here it is convenient to introduce constant $K$ as the coefficient of proportionality in Sievert's law: $K \equiv n_0/\sqrt{P_0}$. It allows us to rewrite Eq. \eqref{s-Sievert} in the following form:
\begin{equation}
\label{s-Sievert-K}
s = K \frac{kT}{\sqrt{\Pi}}.
\end{equation}
Eq. \eqref{s-Sievert-K} shows us that in the case of Sievert's law the solubility value $s$ depends on the final state of the solution. In the current paper a solubility value is understood as gas solubility at the final pressure $\Pi$ -- after the pressure drop.

To deal with the dynamics equation \eqref{Dyn-Eq} one needs to provide it with a reasonable initial condition, i. e. to choose the value of constant $R_i$, initial radius of the bubble, in the following equality
\begin{equation}
\label{R(0)}
R(t)|_{t=0} = R_i.
\end{equation} 
In papers \cite{Epstein-Plesset, Cable-Frade} there was no special meaning assigned to the value of $R_i$: the reason of the appearance of the bubble was totally excluded from discussion. Here we assume that the bubble nucleated fluctuationally, i. e. crossed the barrier corresponding to radius $R_c$. It means that $R_i$ has to exceed $R_c$, but evidently we cannot use the radius of a critical bubble $R_c$ as the initial value for the radius.

A bubble that nucleates fluctuationally in the solution is capable of regular growth if it has passed and, moreover, moved away from the near-critical region where fluctuations are still strong enough. Thus we choose, following Ref. \cite{Kuchma-Gor-Kuni}, $R_i = 2 R_c$. Such value guarantees the absence of fluctuations and, as we will see further, provides us with convenient expressions.

The necessary condition for the fluctuational nucleation of a bubble is a significant pressure drop ($P_0/\Pi \sim 10^3$) and, consequently, due to Eq. \eqref{P_0-zeta}, high supersaturation: $\zeta \sim 30 - 40$. Further we will use the following strong inequality for supersaturation:
\begin{equation}
\label{Strong-Zeta}
\zeta \gg 1.
\end{equation} 

\section{Time dependence of bubble radius}
\label{sec:Radius}

Let us now solve the differential equation \eqref{Dyn-Eq} for the time dependence of the bubble radius with the initial condition \eqref{R(0)} of homogeneous nucleation of the bubble.

The solution of this equation in general case is too cumbersome and even turns out to be unnecessary when inequality \eqref{Strong-Zeta} is fulfilled, as we will ensure in the current section. Here we present solutions of this equation for two particular cases:
\begin{equation}
\label{Small-R}
R \ll \Rs
\end{equation}
and 
\begin{equation}
\label{Large-R}
R \gg R_c.
\end{equation}
When strong inequality \eqref{Strong-Zeta} is fulfilled, these two cases cover the whole range $R \geq 2 R_c$ of regular growth of the bubble radius, and that is the reason why general solution of Eq. \eqref{Dyn-Eq} is not necessary for the system under consideration.

Let us begin with the case when inequality \eqref{Small-R} is fulfilled. In this case we can omit $1$ in comparison with $\Rs/R$ in the l. h. s. of Eq. \eqref{Dyn-Eq}; and we can also omit $1$ in comparison with fraction $3\Rs / 2R$ in the r. h. s. of this equation. Thus we have
\begin{equation}
\label{Growth-Small-R}
\dot{R} = \frac{D s}{\Rs} \left[ (\zeta + 1) - \sqrt{\frac{3}{2} \frac{\Rs}{R}} \right].
\end{equation}
Separating variables and exploiting Eq. \eqref{R_s-R_c} for $\Rs$ under the square root, we can rewrite Eq. \eqref{Growth-Small-R} in the form which make its possible to integrate it
\begin{equation}
\label{Growth-Small-R-1}
\frac{dR}{1 - \sqrt{R_c/R}} = \frac{D s (\zeta + 1)}{\Rs} dt.
\end{equation}
Integrating Eq. \eqref{Growth-Small-R-1} with initial condition \eqref{R(0)} we finally have
\begin{equation}
\label{Growth-Small-R-Final}
R - 2 R_c + R_c \ln \left( \frac{R}{R_c} - 1 \right) + 2 \sqrt{R_c} \left( \sqrt{R} - \sqrt{2 R_c} \right)
 + R_c \ln \left(\frac{\sqrt{2} + 1}{\sqrt{2} - 1} \frac{\sqrt{R} - \sqrt{R_c}}{\sqrt{R} + \sqrt{R_c}} \right)
= \frac{D s}{\Rs}(\zeta + 1) t.
\end{equation}
This formula is different from Eq. (3.4) in Ref.\cite{Kuchma-Gor-Kuni} , which means that, when inequality \eqref{Small-R} is fulfilled, there is a significant difference in the character of growth between Sievert's law and Henry's law.

Now let us proceed to the other case, Eq. \eqref{Large-R}. At first, using Eq. \eqref{R_s-R_c} we can rewrite Eq. \eqref{Dyn-Eq} equivalently in the form of
\begin{equation}
\label{Growth-R_c}
\frac {\dot{R} \left[ R + \Rs \right]}{1 - \sqrt{\frac{1}{(\zeta + 1)^2} + \frac{R_c}{R}}} =D s (\zeta + 1).
\end{equation}
This form makes it obvious that, when both strong inequality \eqref{Large-R} and inequality \eqref{Strong-Zeta} are fulfilled, the whole square root in the denominator of the l. h. s. of Eq. \eqref{Growth-R_c} can be omitted in comparison with $1$. Therefore, we have
\begin{equation}
\label{Growth-Large-R}
\dot{R} \left[ R + \Rs \right] = D s (\zeta + 1).
\end{equation}

\begin{figure}
\includegraphics[height=480pt,angle=270]{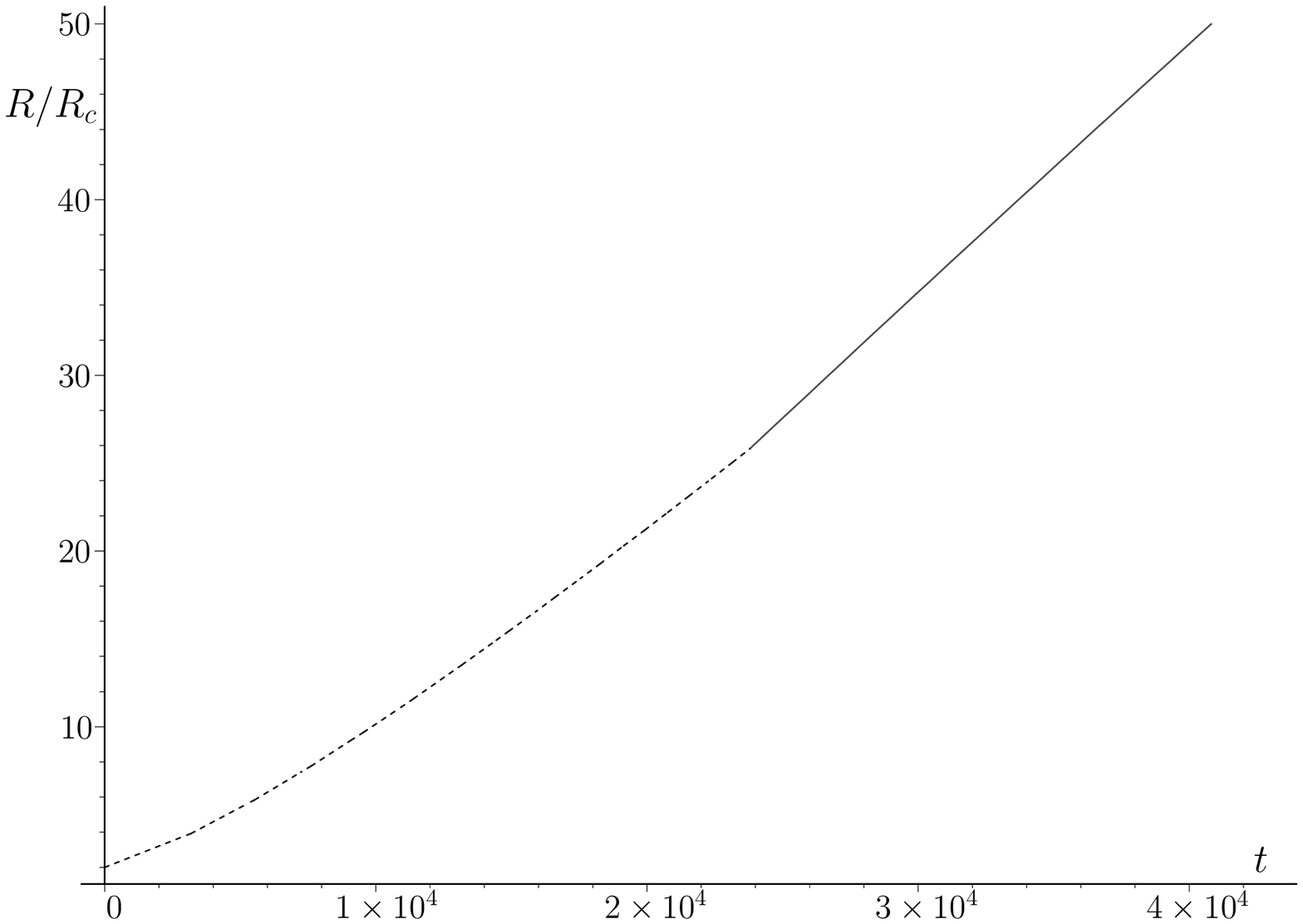}
\caption{Time dependence of bubble radius $R$ from Eq. \eqref{Growth-Small-R-Final} (dotted curve) and Eq. \eqref{Growth-Large-R-2} (solid curve). Time $t$ is measured in units of $\frac{R_c^2}{D s (\zeta + 1)}$.}
\label{Fig:Early}
\end{figure}

This expression can be easily integrated. But initial condition \eqref{R(0)} is not justified when strong inequality \eqref{Large-R} is fulfilled. There is arbitrariness in the choice of the initial condition for integration of Eq. \eqref{Growth-Large-R}, but the most convenient is to choose a condition at such a radius which simultaneously satisfies Eqs. \eqref{Small-R} and \eqref{Large-R}, e. g. 
\begin{equation}
\label{Rm}
\left. R(t) \right|_{t=t_m} = R_m \equiv \sqrt{R_c R_{\sigma}}.
\end{equation}
Due to $R_m \ll R_{\sigma}$, we can use Eq. \eqref{Growth-Small-R-Final} to obtain the explicit value for time $t_m$ defined in Eq. \eqref{Rm}. Integrating Eq. \eqref{Growth-Large-R} with initial condition \eqref{Rm}, and then, substituting the expression for $t_m$, we have
\begin{equation}
\label{Growth-Large-R-2}
\frac{R^2}{2} + \Rs R + 2 R_{\sigma}^{5/4} R_c^{3/4} + R_{\sigma} R_c \left[\ln \left\{ \frac{2^{1/2} + 1}{2^{1/2}  - 1} \left( \left( \frac{R_{\sigma}}{R_c} \right)^{1/4} - 1 \right)^2 \right\} - \frac{5}{2} - 2^{3/2} \right] = D s (\zeta + 1) t.
\end{equation}
Time dependences of bubble radius, when $R \ll \Rs$ (Eqs. \eqref{Growth-Small-R-Final} and \eqref{Growth-Large-R-2}), are presented graphically in Fig. \ref{Fig:Early} for $P_0/\Pi = 10^3$. Bubble radius $R$ is measured in units of $R_c$ and time $t$ is measured in units of $\frac{R_c^2}{D s (\zeta + 1)}$.

With the increase of $R$ the contribution of the third and fourth addends in the l. h. s. of Eq. \eqref{Growth-Large-R-2} decreases; and at $R = R_{\sigma}$ we can already write
\begin{equation}
\label{Growth-Large-R-Final}
\frac{R^2}{2} + \Rs R= D s \zeta t.
\end{equation}
In \eqref{Growth-Large-R-Final} we also took into account strong inequality \eqref{Strong-Zeta}. Eq. \eqref{Growth-Large-R-Final} is identical to Eq. (3.11) in Ref. \cite{Kuchma-Gor-Kuni} . It means that, when inequality $R \geq R_{\sigma}$ is fulfilled, any difference in the character of growth between Sievert's law and Henry's law disappears.

\begin{figure}
\includegraphics[height=480pt,angle=270]{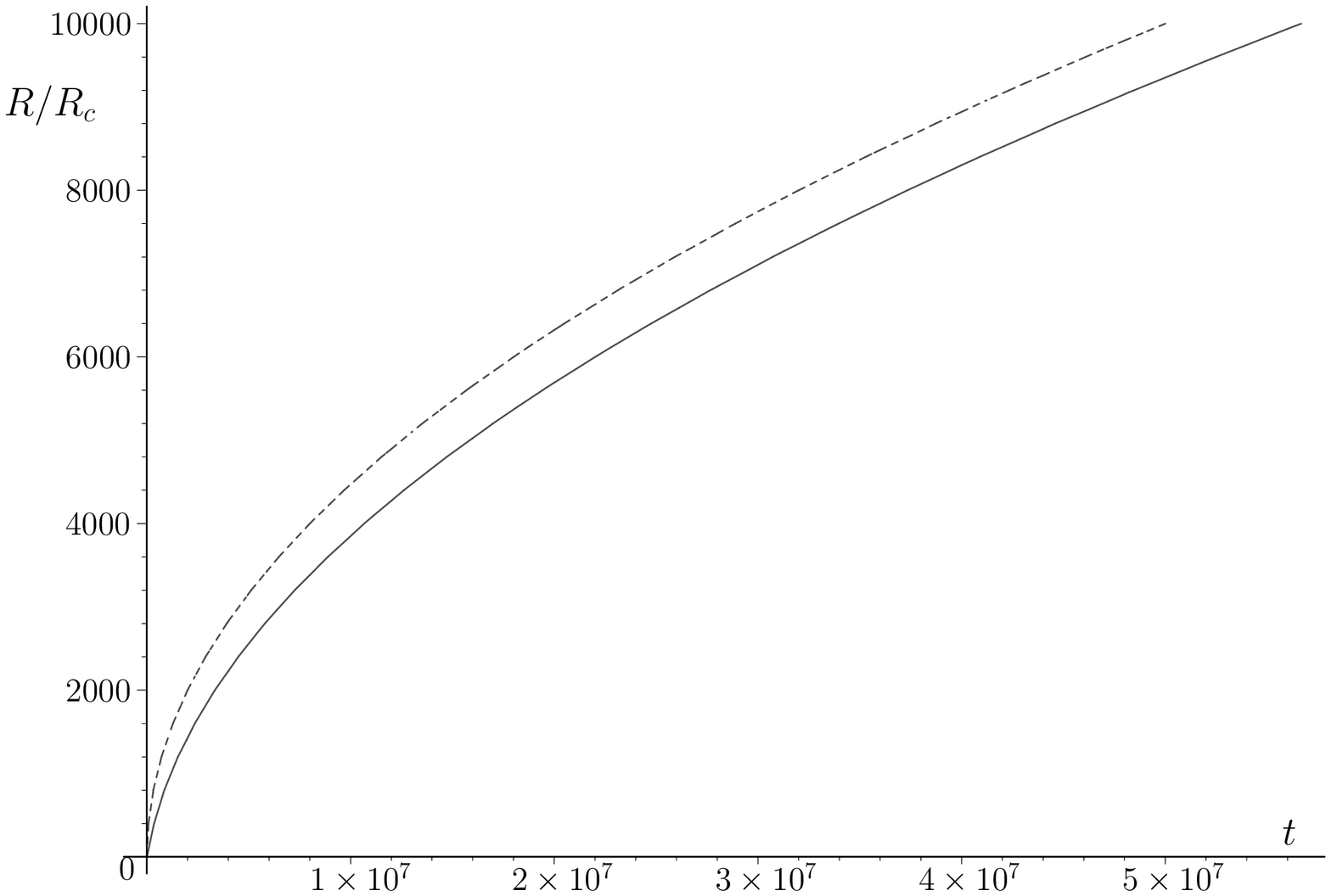}
\caption{Time dependence of bubble radius $R$ from Eq. \eqref{Growth-Large-R-ss} (dotted curve) and Eq. \eqref{Growth-Large-R-2} (solid curve). Time $t$ is measured in units of $\frac{R_c^2}{D s (\zeta + 1)}$.}
\label{Fig:Late}
\end{figure}

When the bubble radius becomes as large as
\begin{equation}
\label{Very-Large-R}
R \gg \Rs,
\end{equation}
Eq. \eqref{Growth-Large-R-Final} transforms into the well-known Scriven's \cite{Scriven-1959} self-similar dependence for the steady-state case
\begin{equation}
\label{Growth-Large-R-ss}
R^2  = 2 D s \zeta t.
\end{equation}
This trend of Eq. \eqref{Growth-Large-R-Final} toward Eq. \eqref{Growth-Large-R-ss} was discussed in detail in Ref. \cite{Kuchma-Gor-Kuni}. This trend is also presented graphically in Fig. \ref{Fig:Late} for $P_0/\Pi = 10^3$. Bubble radius $R$ is measured in units of $R_c$ and time $t$ is measured in units of $\frac{R_c^2}{D s (\zeta + 1)}$.

\section{Conditions for the diffusion flux steadiness}
\label{sec:Stationarity}

\begin{figure}
\includegraphics[height=480pt,angle=270]{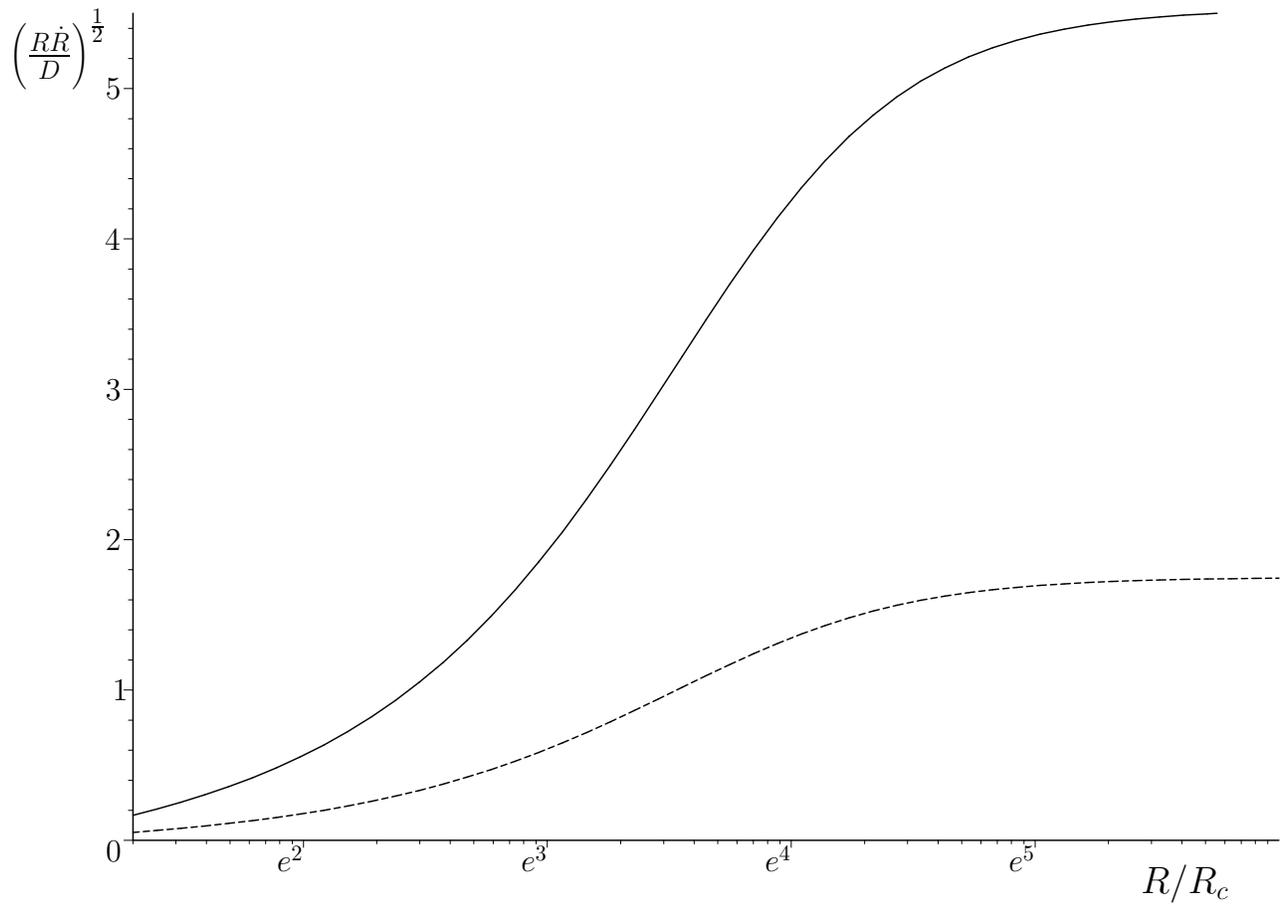}
\caption{Parameter characterizing the steadiness of bubble growth as a function of bubble radius $R$. Solid curve: $s = 1$, dotted curve: $s = 10^{-1}$.}
\label{Fig:Steadiness}
\end{figure}

The diffusion flux of molecules toward a growing bubble can be considered steady when the bubble growth is slow enough in comparison with the ``diffusion cloud'' growth. To be more exact, the radius of the bubble has to be much smaller than the radius of this cloud -- the diffusion length. We can express it as
\begin{equation}
\label{R<<l}
R \ll (D t_R)^{1/2},
\end{equation}
where $t_R$ is the characteristic time of the bubble radius change, $t_{R} \equiv R/\dot{R}$ is the time in which the bubble radius changes significantly. Evidently, Eq. \eqref{R<<l} can be rewritten as
\begin{equation}
\label{RR<<1}
\left( R \dot{R} / D \right)^{1/2} \ll 1.
\end{equation}
We can make this condition more explicit by means of Eq. \eqref{Dyn-Eq}
\begin{equation}
\label{Steady-General}
\left( s \frac{(\zeta + 1) - \sqrt{1 + \frac{3}{2} \frac{\Rs}{R}}}{1 + \frac{\Rs}{R}} \right)^{1/2} \ll 1.
\end{equation}
It should be noted that condition \eqref{Steady-General} is valid regardless of whether the bubble nucleation was homogeneous or not (the same can be applied to Eq. \eqref{Dyn-Eq}). Graphical dependence of parameter characterizing the steadiness of bubble growth (l. h. s. of Eq. \eqref{Steady-General}) on bubble radius $R$ is presented in Fig. \ref{Fig:Steadiness}. In Fig. \ref{Fig:Steadiness} bubble radius is measured in units of $R_c$ and presented in logarithmic scale, $P_0/\Pi = 10^3$, solid curve: $s = 1$, dotted curve: $s = 10^{-1}$.

Now, exploiting Eq. \eqref{Steady-General}, let us obtain conditions of steadiness of the diffusion flux at different characteristic sizes that we have for the case of homogeneous nucleation of the bubble. Obviously, the larger the bubble is, the stricter the condition for steadiness becomes (see Fig. \ref{Fig:Steadiness}). The general condition of the steadiness of bubble growth at any time is the condition at $R \gg \Rs$. From Eq. \eqref{Steady-General} we have
\begin{equation}
\label{Steady-3}
s^{1/2} \ll \left( \frac{1}{\zeta} \right)^{1/2}.
\end{equation}
As we will find in the following section, Eq. \eqref{Steady-3} is difficult to satisfy. But it is evident that in order to satisfy this condition one needs to decrease the solution supersaturation $\zeta$ (the case of heterogeneous nucleation) or, leaving supersaturation $\zeta$ constant, decrease the gas solubility $s$. From Eq. \eqref{P_0-zeta} for the supersaturation we have
\begin{equation}
\label{zeta-Sievert}
\zeta = \sqrt{\frac{P_0}{\Pi}} - 1.
\end{equation}
Then, using Eqs. \eqref{s-Sievert-K}, \eqref{zeta-Sievert}, let us rewrite product $s \zeta$ in the following form
\begin{equation}
\label{s-zeta}
s \zeta = K \frac{kT}{\sqrt{\Pi}} \left( \sqrt{\frac{P_0}{\Pi}} - 1 \right).
\end{equation}
Eq. \eqref{s-zeta} shows us that in order to weaken limitation \eqref{Steady-3} one needs to increase final pressure $\Pi$, leaving ratio $P_0/\Pi$ constant.

The condition of steadiness at the very beginning of the regular growth of the bubble, can be found, exploiting Eq. \eqref{Steady-General} with $R = 2 R_c$:
\begin{equation}
\label{Steady-0}
s^{1/2} \ll \zeta^{1/2}.
\end{equation}
While deriving Eq. \eqref{Steady-0}, the numerical coefficient $\left( 3 \left( 1 - 1/\sqrt{2} \right) \right)^{1/2}$ in its l. h. s. was replaced with $1$ for shortness.

\section{Bubble growth in magma during volcano eruptions}
\label{sec:Volcanoes}

This section contains the analysis of the question whether the steady growth conditions obtained in section \ref{sec:Stationarity} are fulfilled for bubble nucleation in volcanic systems. Previously \cite{Navon,Chernov-2004} the study of such systems exploited steady approximation without the analysis of its applicability.

It should be noted that usually there are several different gases dissolved in magma (e.~g. $HCl$, $HF$, $H_2S$, $SO_2$, $H_2$ and $CO_2$), but the principal component is still water vapor \cite{Burnham} . For the analysis of steady conditions discussed below the presence of other gases is not important; therefore, we will consider the case of homogeneous nucleation of water vapor bubbles in magma.

It is well-known that Sievert's law is fulfilled for gases where diatomic molecules dissociate during gas dissolution \cite{Landau-V} . While water has a triatomic molecule, its vapor in magma obeys Sievert's law. This happens because each water molecule together with oxygen from silicate melt forms two hydroxyl groups (see Eq. (2) in Ref. \cite{Stolper}):
\begin{equation}
\label{Stolper}
H_2O + O = 2 OH.
\end{equation}

For our evaluations we will use parameters from Ref. \cite{Chernov-2004} ; we have 
\begin{equation}
\label{data}
P_0 = 100~\text{MPa} ~~~ \Pi = 0,1~\text{MPa} ~~~ T = 1150~\text{K} ~~~ w = 3\% ~~~ \rho_m = 2300~\text{kg}/\text{m}^3.
\end{equation}
Here $w$ is gas mass fraction of the dissolved gas (water vapor), and $\rho_m$ is magma density. Notwithstanding the fact that liquid magmatic melt is a solution of several gases, none of these gases are surfactants and the solution is diluted; therefore, we can characterize such solution with surface tension $\sigma$. Navon et al.\cite{Navon} gives the following values for the surface tension of magmatic melts: $\sigma \sim (0.05 - 0.1)~\text{N}/\text{m}$.

Let us express the values of $s$ and $\zeta$ using data given. From Eq. \eqref{zeta-Sievert}, in accordance with Eq. \eqref{data}, we have $\zeta \simeq 31$. Then we need to calculate $n_0$ and substitute it into Eq. \eqref{s-Sievert} to obtain the value of solubility. As long as we are given mass density of magma and mass fraction of gas, it is convenient to write
\begin{equation}
\label{n_0-rho_0}
n_0 = \rho_0 \frac{N_A}{\mu},
\end{equation}
where $\rho_0$ is mass density of the dissolved gas, $N_A = 6 \times 10^{23}~\text{mol}^{-1}$ is the Avogadro constant and $\mu = 1.8 \times 10^{-2}~\text{kg}/\text{mol}$ is the molar mass of the dissolved gas (water). Finally, we need to express mass density of the dissolved gas. Evidently, we have
\begin{equation}
\label{rho_0}
\rho_0 = w \rho_m,
\end{equation}
and, therefore, 
\begin{equation}
\label{s-Chernov}
s = w \rho_m \frac{N_A k T}{\mu \sqrt{P_0 \Pi}}.
\end{equation}
Using data \eqref{data} in Eq. \eqref{s-Chernov}, we have $s \simeq 12$.

Now we can see that for volcanic systems, where the pressure drop is of the order of $10^3$ and solubility is more than $1$, condition \eqref{Steady-3} is violated; and steady approximation is not valid for radii of the order of $R_{\sigma}$. Even in the very beginning of bubble regular growth, when $R = 2 R_c$, the steady condition \eqref{Steady-0} is fulfilled only at its breaking point: the values of $\zeta^{1/2}$ exceed the value of $s^{1/2}$, but these values are of the same order of magnitude.

Let us make the following interesting observation. Since for $R \gg R_{\sigma}$ the bubble dynamics is identical for both Henry's and Sievert's solubility laws, the condition of steady growth for $R \gg R_{\sigma}$ is also the same (see Eq. \eqref{Steady-3} above and Eq. (2.14) in Ref. \cite{Kuchma-Gor-Kuni}). For the case of homogeneous nucleation, when the pressure drop $P_0/\Pi \sim 10^3$, the steady condition at $R \gg R_{\sigma}$ is, as a rule, violated in cases of both Henry's and Sievert's laws. For Henry's law, when $s \sim 10^{-2}$, it is violated due to high supersaturation values $\zeta \sim 10^{3}$. For Sievert's law, when, due to $P_0/\Pi \sim 10^3$ and Eq. \eqref{P_0-zeta}, corresponding supersaturation values are significantly less $\zeta \sim 30 - 40$, values of gas solubility are significantly higher than for Henry's law, and that is the reason of the violation of the steady condition.

\section{Conclusions}
\label{sec:Conclusions}

In the presented paper we obtained the equation to a bubble growth dynamics in the supersaturated gas solution with Sievert's solubility law. We showed that, as long as the bubble size is such that Laplace forces influence the bubble growth, the difference between Henry's and Sievert's laws in terms of chemical equilibrium conditions results in the difference in bubble growth dynamics. When Laplace forces become weaker, the bubble growth dynamics is identical in the cases of both Henry's and Sievert's laws.

While obtaining the dynamics equation we assumed the diffusion flux to be steady. We obtained conditions when this steady approximation is applicable. We showed that usually, as the radius of the bubble increases, the steady regime of bubble growth gradually gives way to the nonsteady one. Application of the obtained conditions for the volcanic system consisting of water vapor dissolved in silicate melt showed that the process in such system cannot be considered as steady.

\section*{Acknowledgments}
\label{sec:Acknowledgments}
The research has been carried out with the financial support of the Russian Analytical Program "The Development of Scientific Potential of Higher Education" (2009-2010): project RNP.2.1.1.4430. "Structure, Thermodynamics and Kinetics of Supramolecular Systems". One of the authors (G.~G.) is also grateful to the K.~I.~Zamaraev foundation for the support of his post-graduate research. 

\newpage 


\newpage

\end{document}